\documentclass[prd,amsfonts,onecolumn,superscriptaddress,aps,nofootinbib,11pt]{revtex4-2}

\usepackage[english]{babel}
\usepackage[utf8]{inputenc}

\usepackage[pdftex, pdftitle={Vacuum Stability in the one-loop approximation of a 331 Model}, colorlinks,citecolor=blue,urlcolor=blue,linkcolor=blue]{hyperref}
\usepackage{amsthm}
\usepackage{mathtools}
\usepackage{physics}
\usepackage{xcolor}
\usepackage{graphicx}
\usepackage[top=3cm,bottom=3cm,left=2cm,right=2cm,marginparwidth=1.75cm]{geometry}
\usepackage{adjustbox}
\usepackage{placeins}
\usepackage[T1]{fontenc}
\usepackage{csquotes}
\usepackage{array,multirow}
\usepackage{color}
\usepackage{bm} 
\usepackage{inputenc}
\usepackage{subfigure}

\providecommand{\ZZ}{\mathbb{Z}}

\begin{document}
\title{Vacuum Stability in the one-loop approximation of a 331 Model}

\author{G. C. Dorsch}
    \email[Correspondence email address: ]{glauber@fisica.ufmg.br}
\author{A. A. Louzi}
    \email[Correspondence email address: ]{alvarolouzi@ufmg.br} 
\author{B. L. Sánchez-Vega}
    \email[Correspondence email address: ]{bruce@fisica.ufmg.br}
\author{A. Viglioni}
    \email[Correspondence email address: ]{arthurcesar@ufmg.br}   
    \affiliation{Departamento de F\'isica, UFMG, Belo Horizonte, MG 31270-901, Brasil.\label{addr1}}
\begin{abstract}

In this study, we analyze the vacuum stability of the economical 331 model at the one-loop level using the renormalization group equations and a single-scale renormalization method. By integrating these equations, we determine stability conditions up to the Planck scale, incorporating constraints from recent experimental data on new Higgs-like bosons, charged scalars, and charged and neutral gauge bosons. Our analysis uncovers intriguing relations between the mass of the heaviest scalar and the masses of exotic quarks, in order to ensure stability of the model up to the Planck scale. For the 331 energy scale used in this work, $18$ TeV, we find an upper bound on the heaviest quark mass of the model, which is not so distant from future LHC runs, serving as bounds to be searched.  Additionally, we explore relations between the scalar couplings coming stability and perturbativity conditions. These impose unprecedented constraints on the economical 331 model.
\end{abstract}

\maketitle


\section{Introduction \label{introduction}}
It is widely recognized that the Standard Model (SM) is unable to provide explanations to some observed phenomena, such as the extensive evidence supporting the existence of dark matter \cite{ExistenceDarkMatter,PlanckDM2020}, the neutrinos masses and their mixing \cite{Yosef-Garcia2003,NeutrinoOrigin,Neutrino3,NeutrinoDM,Neutrino1,Neutrino2}, and baryonic asymmetry \cite{shaposhnikov1987baryon,kitano2005dark}. In addition to these issues, the recent measurement of the Higgs mass \cite{ATLASHiggs,CMSHiggs} places the SM near the metastability and stability boundaries \cite{Degrassi_2012,Buttazzo_2013,Andreassen_2014}. In these studies, the SM is extrapolated to the Planck scale using the two-loop effective potential to analyze the quartic coupling and its beta function in the proximity of this high-energy regime. For the Higgs mass value of $m_h=125.25\pm0.17$ GeV \cite{ATLAS:2015yey,ATLAS:2018tdk,CMS:2020xrn}, the authors identified an instability scale significantly below Planckian physics, approximately at $10^{11}$~GeV. According to their interpretation, this finding could be an indicative of the presence of new physics around this scale.

In the context of the SM, classical stability is mostly examined through the quartic terms of the Higgs field, $h$, which for very large values of $h$ takes the form 
\begin{equation*}
    V(h)\approx \frac{\lambda}{4}h^4.
\end{equation*}

Therefore, in this form, the classical potential is stable (bounded from below, BFB) if $\lambda > 0$. As shown in ref. \cite{sher1989electroweak}, $\lambda$ could not be greater than $4\pi$, since perturbative theory would break down, nor be very small, since radiative corrections become important. However, the classical potential is found to be insufficient for analyzing the stability of the SM across diverse energy scales since quantum corrections are typically relevant. To address this limitation, the effective potential is employed. As pointed out in ref. \cite{Casas_1995}, one can always define an effective potential for the SM up to two loops such that, for $h\gg v$, it assumes the form
\begin{equation*}
    V_{\text{eff.}}(h)=\frac{\lambda_{\text{eff}}(h)}{4}h^4.
\end{equation*}
This approximation is valid as the instability scale occurs at energies much larger than the electroweak scale. Given that the effective potential shares the same form as the classical potential, one concludes the effective potential stability is realized for $\lambda_{\text{eff}}>0$.  In conclusion, the examination of the stability of the SM is comparatively straightforward, given that both the classical and effective potentials, up to the second loop order, must satisfy an explicit condition (namely $\lambda>0$ or $\lambda_\text{eff}>0$) that are easily fulfilled.

On the other hand, models beyond Standard Model (BSM) exhibit a more intricate potential~\cite{Pisano:1992bxx,Foot:1992rh,Dias:2003zt,ellwanger2010next,branco2012theory,Bonilla_2015,Dias_2020,Dias:2022hbu}. Typically, these models contain additional scalar fields that interact among themselves increasing the amount of quartic terms, rendering it difficult to determine classical stability conditions, as well as the analytical mass eigenvalues. The situation is aggravated when attempting to study the corresponding effective potential. The logarithm terms coming from the one-loop corrections depend on ratios of mass eigenvalues, obtained from the Hessian matrix of the tree level potential, and the renormalization scale. In a situation where the mass values are very different, it is not possible to define a single renormalization scale that maintains all logarithmic terms under perturbative control. Such effective potentials are referred to in the literature as multiscale \cite{abud1981geometry,nie1999vacuum,arhrib2011higgs,kannike2012vacuum,kannike2016vacuum}.

The determination of classical stability conditions for multiscale potentials has been extensively explored, as documented in refs. \cite{abud1981geometry,nie1999vacuum,arhrib2011higgs,kannike2012vacuum,kannike2016vacuum}. Generally, for analytical determination, a simplifying condition is commonly imposed to ensure that the quartic terms of the scalar potential take the form $\lambda_{ij}\phi^2_{i}\phi^2_{j}$, making them essentially biquadratic. Therefore, requiring that the scalar potential be bounded from below as the fields approach infinity amounts to ensuring that the matrix $\lambda_{ij}$ is copositive (conditionally positive). Furthermore, the use of the orbital space method~\cite{kim1982general} is a common practice to reduce the number of variables, thus increasing the tractability of the problem.

For the multiscale effective potential, some methods were employed in refs.~\cite{einhorn1984new,ford1993effective,bando1993improving,ford1994multiscale,steele2014multiscale}. These methods introduce different scales of renormalization and, therefore, one must work with partial renormalization group equations. Due to this technical complication, the authors in ref.~\cite{chataignier2018single} developed a method to improve the effective potential for any order, using the renormalization group equations with a single renormalization scale. Notably, it has been applied to analyze the extremum structure of scale-invariant effective potentials in a previous work~\cite{Kannike:2020ppf}. An interesting feature of this method is that it allows assessment of improved effective potential using tree-level potential, provided that the parameters remain perturbative. The method involves solving the renormalization group equations for the effective potential, choosing the boundary condition on the hypersurface so that the quantum corrections disappear. Thus, tree-level stability criteria are applicable to the improved effective potential. In other words, it allows us to replace the quartic couplings under classical tree-level stability conditions by the running couplings obtained from solving the equations of the renormalization group. Therefore, it is possible to study the quantum stability conditions for any model, using the copositivity criteria and the single scale renormalization to multiscale effective potential.
\par
We are interested in applying these methods to the economical 331 model~\cite{foot19943,ponce2002analysis}, belonging to a class of models known as 331 models. The name arises as an abbreviation of the semi-simple gauge group $\text{SU(3)}_C\times\text{SU(3)}_L\times\text{U(1)}_N$. The electric charge operator is defined as $Q=T_3-bT_8+N\mathbf{1}_{3\times3}$, where $T_{3,8}$ are the  $\text{SU}(3)_L$ diagonal generators and $N$ is the charge associated with the abelian group U(1). The allowed values for $b$ are $\sqrt{3}$ or $1/\sqrt{3}$. The  economical 331 model belongs to the latter value of $b$. This model, when compared to others 331 models, as in refs. \cite{Pisano:1992bxx,Foot:1992rh}, can accommodate a simple scalar structure consisting of three scalar triplets $\chi,\rho,\eta$ without exotic electric charge. In addition, it is also able to accommodate a right-handed neutrino in the leptonic triplet since the $b=1/\sqrt{3}$ value allows two electrically neutral components. This property allows massive neutrinos at tree level, although quantum corrections are necessary to make them agree with the experiments data~\cite{Dong_2007}. Another interesting feature is the implementation of the Peccei-Quinn mechanism in order to solve strong CP-problem \cite{PQMec} and the existence of axion dark matter \cite{DarkMatter1,AxionDM1}. 
\par
Our objective is to extend the investigation into vacuum stability initiated in ref. \cite{S_nchez_Vega_2019}. In that work, the authors introduced a $\ZZ_2$ discrete symmetry to the model, facilitating the expression of the classical quartic scalar potential in a biquadratic form. This formulation enables the application of copositivity criteria \cite{kim1982general,kannike2012vacuum,kannike2016vacuum}, leading to the derivation of seventeen classical stability conditions. Solving these conditions, in conjunction with constraints derived from the first and second derivative tests for the potential minima, has, for the first time, allowed the establishment of analytical constraints for the coupling constants.

In order to improve these constrains, a quantum analysis of the vacuum stability within the 331 economical model becomes imperative. In our approach, we employ the previously mentioned single-scale renormalization method~\cite{chataignier2018single} to address the effective potential in the context of the renormalization group equations. Given the elevated complexity of the model, we employ the Mathematica RGBeta package, as outlined in ref.~\cite{thomsen2021introducing}, to compute all required beta functions. Subsequently, we numerically solve the eighteen coupled nonlinear differential equations. This process is essential for identifying the new allowed region for the quartic couplings, ensuring that the effective potential remains bounded from below to the Planck scale. As elaborated in Sec. \ref{stability}, our findings reveal a significantly diminished region for the allowed values of the coupling constants in comparison to the outcomes presented in ref.~\cite{S_nchez_Vega_2019}. Moreover, we will explore the interplay among the masses of the exotic quarks and the heaviest scalar.

The rest of the paper is organized as follows. In Sec. \ref{model} we present the generalities of the model with an imposed $\ZZ_2$ symmetry, bringing simplicity and allowing the scalar potential to be biquadratic in the fields norms. In Sec. \ref{mass}, we present the mass spectrum of the model in order to determine or constrain the couplings, such as quartic, gauge and Yukawa couplings. Specifically, the expressions for the CP-even masses  will be important to study the stability of the model. Moreover, by incorporating complementary information from the $Z-Z'$ mixing angle, the lower bound on a new charged gauge boson, and the hierarchy of vacuum expectation values in the model, we have determined the minimum value of the first symmetry breaking parameter, $v_{\chi}$. In Sec. \ref{stability}, by applying the tree level stability criteria in the improved effective potential, we obtain interesting regions on the quark-Higgs plane and, in the quartic couplings parameter space. Finally, we present our conclusions in Sec. \ref{conclusions}.


\section{The generalities of the model \label{model}} 
An appealing aspect of 331 models is the partial explanation provided for the number of families, a feature absent in the SM. This distinction arises from the fact that, in the SM, the cancellation of gauge anomalies takes place generation by generation. In contrast, in 331 models, the condition for anomaly cancellation implies that the number of families is either three or a multiple of three.\par
In the present context of the economical 331 model, requiring absence of gauge anomaly dictates that the fermionic field content must be
\begin{align}
\textrm{Leptons: }
L_{aL}&=\left(\nu_{a},\,e_{a},\,N_{a}^{c}\right)_{L}^{\textrm{T}}\sim\left(1,\mathbf{3},\,-1/3\right), \label{eq:1}\\
\textrm{Quarks: }Q_{L} & =\left(u_{1}\nonumber,\,d_{1},\,u{}_{4}\right)_{L}^{\textrm{T}}\sim\left(\mathbf{3},\,\mathbf{3},\,1/3\right),\\
Q_{bL} & =\left(d_{b},\,u_{b},\,d_{b+2}\right)_{L}^{\textrm{T}}\sim\left(\mathbf{3},\,\bar{\mathbf{3}},\,0\right),\label{eq:2}
\end{align}
for the left-handed fields, whereas for the right-handed fields
\begin{align}
\textrm{Leptons: }e_{aR} & \sim\left(1,\,1,\,-1\right),\label{eq:3}\\
\textrm{Quarks:\,\ }u_{sR} & \sim\left(3,\,1,\,2/3\right),\quad d_{tR}\sim\left(3,\,1,\,-1/3\right),\label{eq:4}
\end{align}
where $a=1,2,3$, $b=2,\,3$, $s=1,\dots,4$, $t=1,\dots,5$ and ``$\sim$'' means the transformation properties under the local gauge group.\par
In order to completely break the gauge symmetry down to $\textrm{U}\left(1\right)_{Q}$ and, at the same time, to give the phenomenologically appropriate masses for all particles, it is necessary to include three scalar triplets in the fundamental representation~\cite{Montero_2015},
\begin{equation}
\label{eq:5}
\rho=\left(\rho_{1}^{+},\,\rho_{2}^{0},\,\rho_{3}^{+}\right)^{\textrm{T}}\sim\left(1,\,\mathbf{3},\,2/3\right), \quad\eta=\left(\eta_{1}^{0},\,\eta_{2}^{-},\,\eta_{3}^{0}\right)^{\textrm{T}}\sim\left(1,\:\mathbf{3},\,-1/3\right),
\end{equation} 
\begin{equation}
\label{eq:6}
\quad\chi=\left(\chi_{1}^{0},\,\chi_{2}^{-},\,\chi_{3}^{0}\right)^{\textrm{T}}\sim\left(1,\:\mathbf{3},\,-1/3\right).
\end{equation}
The introduction of a $\mathbb{Z}_2$ discrete symmetry, acting on fields as $\chi\rightarrow-\chi$, $u_{4R}\rightarrow-u_{4R}$, and $d_{\left(4,5\right) R }\rightarrow-d_{\left(4,5\right)R}$, while leaving the other fields unaffected, not only simplifies the model but also mitigates FCNC problems \cite{DUMM_1994}. Additionally, it facilitates the implementation of the PQ mechanism \cite{PQMec} and, in certain scenarios, contributes to the stabilization of potential dark matter candidates \cite{331DM,DarkMatter2,DarkMatter1}.\par
Taking into consideration all the fermionic and bosonic fields and their symmetries, the most general renormalizable Yukawa Lagrangian reads
\begin{equation}
\label{eq:7}
\mathcal{L}_{\textrm{Yuk}}=\mathcal{L}^{\rho}+\mathcal{L}^{\eta}+\mathcal{L}^{\chi},
\end{equation}
with
\begin{eqnarray} 
-\mathcal{L}^{\rho} & = & \alpha_{a}\bar{Q}_{L}d_{aR}\rho+\alpha_{ba}\bar{Q}_{bL}u_{aR}\rho^{*}+\text{Y}_{aa'}\varepsilon_{ijk}\left(\bar{L}_{aL}\right)_{i}\left(L_{a'L}\right)_{j}^{c}\left(\rho^{*}\right)_{k}+\text{Y}'_{aa'}\bar{L}_{aL}e_{a'R}\rho+\nonumber \\
 &  & \textrm{H.c.,}\label{eq:8}\\
-\mathcal{L}^{\eta} & = & \beta_{a}\bar{Q}_{L}u_{aR}\eta+\beta{}_{ba}\bar{Q}_{bL}d_{aR}\eta^{*}+\textrm{H.c.},\label{eq:9}\\
-\mathcal{L}^{\chi} & = & \gamma_{4}\bar{Q}_{L}u_{4R}\chi+\gamma{}_{b\left(b+2\right)}\bar{Q}_{bL}d_{\left(b+2\right)R}\chi^{*}+\textrm{H.c.}\,,\label{eq:10}
\end{eqnarray}
where $\epsilon_{ijk}$ is the Levi-Civita symbol and $a^\prime,i,j,k=1,2,3$ and $a$, $b$, $s$, $t$ are in the same range as in eqs.~(\ref{eq:1}-\ref{eq:4}). Additionally, it is noteworthy that the $u_{4R}$ and $d_{(4,5)R}$ quarks exclusively couple to the $\chi$ triplet. This distinction allows us to identify the $\chi$ field as the  responsible for the initial spontaneous symmetry breaking, whereas $\eta$ and $\rho$ will play a role in the subsequent breaking, as we will soon elucidate.\par
In addition, the most general and renormalizable scalar potential invariant under the gauge and $\ZZ_2$ symmetries is  
\begin{eqnarray}
\label{eq:11}
V\left(\eta,\rho,\chi\right)& = & -\mu_{1}^{2}\eta^{\dagger}\eta-\mu_{2}^{2}\rho^{\dagger}\rho-\mu_{3}^{2}\chi^{\dagger}\chi \nonumber\\
 &  & +\lambda_{1}\left(\eta^{\dagger}\eta\right)^{2}+\lambda_{2}\left(\rho^{\dagger}\rho\right)^{2}+\lambda_{3}\left(\chi^{\dagger}\chi\right)^{2}+\lambda_{4}\left(\chi^{\dagger}\chi\right)\left(\eta^{\dagger}\eta\right)\nonumber \\
 &  & +\lambda_{5}\left(\chi^{\dagger}\chi\right)\left(\rho^{\dagger}\rho\right)+\lambda_{6}\left(\eta^{\dagger}\eta\right)\left(\rho^{\dagger}\rho\right)+\lambda_{7}\left(\chi^{\dagger}\eta\right)\left(\eta^{\dagger}\chi\right)\nonumber \\
 &  & +\lambda_{8}\left(\chi^{\dagger}\rho\right)\left(\rho^{\dagger}\chi\right)+\lambda_{9}\left(\eta^{\dagger}\rho\right)\left(\rho^{\dagger}\eta\right)+\lambda_{10}\left(\chi^{\dagger}\eta\right)^{2}\nonumber\\
 &  & -\frac{\lambda_{15}}{\sqrt{2}}\epsilon_{ijk}\eta_{i}\rho_{j}\chi_{k}+\textrm{H.c.}\,.
\end{eqnarray}
Note that the term $\lambda_{15}\epsilon_{ijk}\eta_{i}\rho_{j}\chi_{k}$ effectively breaks the $\ZZ_2$ symmetry, but must be included. Its absence would lead to the emergence of a QCD axion with a small decay constant, $11.5\text{ keV}\leq f_{a}\leq 246\text{ GeV}$ \cite{Sanchez-Vega:2016dwe}, already ruled out by experiments \cite{Bardeen:1986yb}. Furthermore, we use the freedom of definition of scalar fields to make $\lambda_{10}$ and $\lambda_{15}$ positive real numbers.


\section{Mass spectrum \label{mass}}
To study the vacuum stability at the one-loop level, we must first determine the mass spectrum. For this reason, in this section we find the analytical expression, at tree level, for particle masses in the model. The minimal conditions for giving mass to all particles is that the scalar fields gain the following vacuum expectation values (VEVs)
\begin{equation}\label{eq:12}
    \langle\rho\rangle=\frac{1}{\sqrt{2}}(0,\,v_{\rho},\,0)^{\text{T}},\ 
    \langle\eta\rangle=\frac{1}{\sqrt{2}}(v_{\eta},\,0,\,0)^{\text{T}},\ 
    \langle\chi\rangle=\frac{1}{\sqrt{2}}(0,\,0,\,v_{\chi})^{\text{T}}.
\end{equation}
The first step of the spontaneous symmetry breaking is done by $v_\chi$, whilst the second is realized by $v_\rho$ and $v_\eta$, that is,
\begin{equation*}
\text{SU(3)}_L\otimes\text{U(1)}_N
\xrightarrow{v_\chi}\text{SU(2)}_L\otimes\text{U(1)}_Y\xrightarrow{v_\eta, \, v_\rho}\text{U(1)}_Q.
\end{equation*}
The starting point of our analysis will be the gauge sector, because, besides being the simplest, it will allow us to estimate VEVs and gauge couplings. Following standard procedures, we define the covariant derivatives as $D_{\mu}=\partial_{\mu}-ig_L\,W^{a}_{\mu}\,\lambda^{a}/2-ig_NN\,B_{\mu}$, where $\lambda^{a}$ are the Gell-Mann matrices and $g_L$ and $g_N$ are the gauge couplings of the SU$(3)_L$ and U$(1)_N$ groups, respectively. After some straighforward calculations, we find that the non-hermitian charged gauge bosons and their respective masses are \footnote{Note the different signs appearing in the definitions of $Y^\pm$ and $W^\pm$.}
\begin{eqnarray}
    &&W^{\pm}_{\mu}=\frac{1}{\sqrt{2}}(W^1_{\mu}\mp iW^2_{\mu}),\quad X^{0(*)}_{\mu}=\frac{1}{\sqrt{2}}(W^4_{\mu}\mp iW^5_{\mu}),\quad Y^{\pm}_{\mu}=\frac{1}{\sqrt{2}}(W^6_{\mu}\pm iW^7_{\mu}), \quad \label{eq:13}\\
     &&M^2_{W^{\pm}}=\frac{g_L^2}{4}(v_{\eta}^2+v_{\rho}^2),\quad\ \  M^2_{X^{0(*)}}=\frac{g_L^2}{4}(v_{\eta}^2+v_{\chi}^2),\quad\ \ \ \  M^2_{Y^{\pm}}=\frac{g_L^2}{4}(v_{\rho}^2+v_{\chi}^2). \label{eq:14}
\end{eqnarray}
Note that the new gauge bosons $X^{0(*)}_{\mu}$ and $Y^{\pm}_{\mu}$ gain masses at the first step of spontaneous symmetry breaking, and a mass correction coming from the second step. From the experimental value for the SM $W$ gauge boson, $M_{W^{\pm}}=80.377\pm0.012$ GeV \cite{MW1,MW2}, we calculate that $v^2_{\text{SM}}\equiv v_{\eta}^2+v_{\rho}^2=(246~\text{GeV})^2$ and $g_L\simeq0.66$.

\begin{figure}[h!]
    \centering
    \includegraphics[scale=0.35]{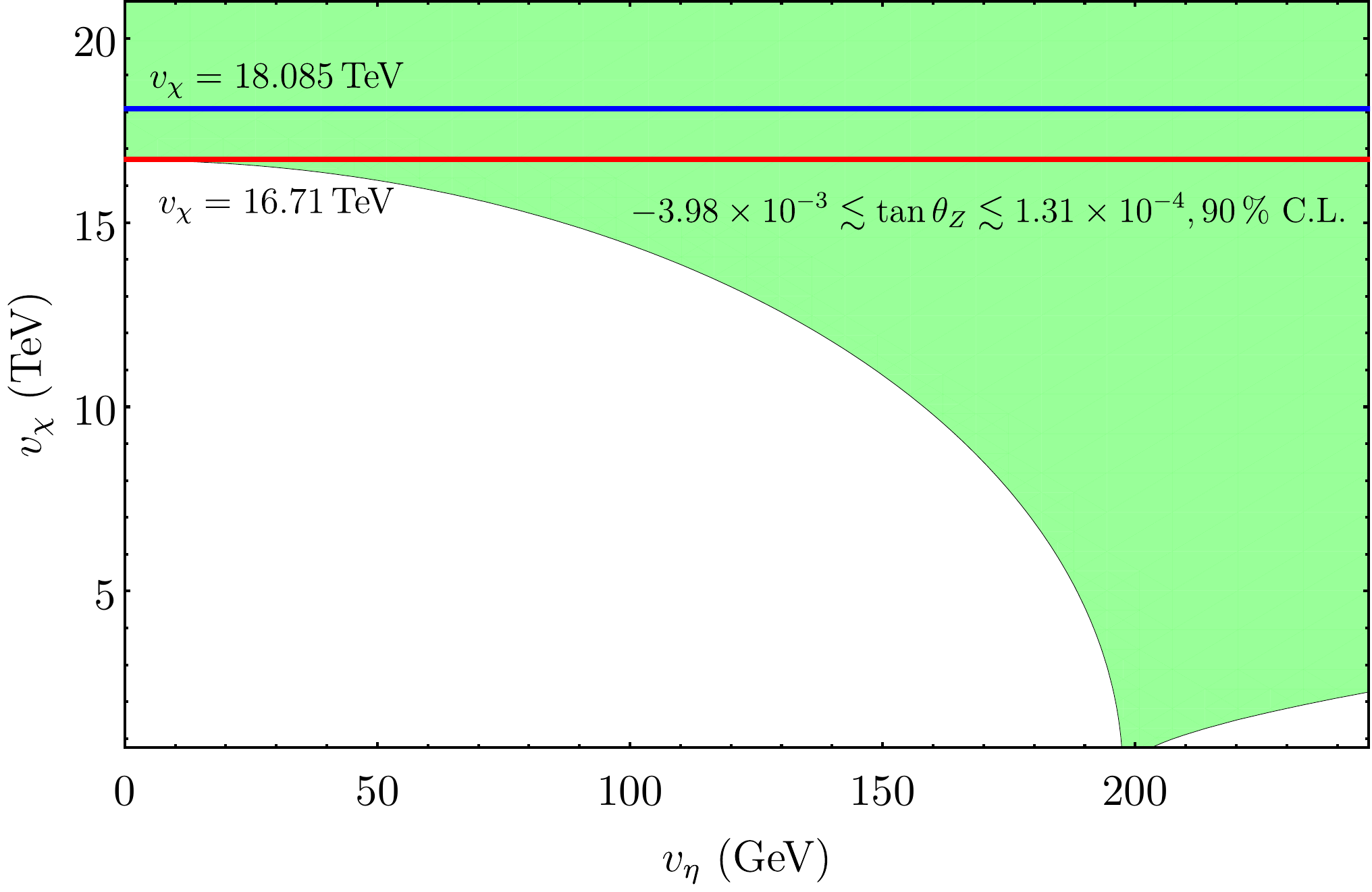}
    \caption{The allowed values for $v_{\eta}$ and $v_{\chi}$ are determined by considering $M^2_{ZZ'},\,M^2_{Z}$, and $M^2_{Z'}$ as functions of these VEVs. It is important to note that, for values of $v_{\chi}\gtrsim$ 16.71 TeV, there are no restrictions on the values of $v_{\eta}$.}
    \label{fig:veta-vchi}
\end{figure}

Note that to determine the masses of the $X^{0(*)}_{\mu}$ and $Y^{\pm}_{\mu}$ gauge bosons, knowledge of the value of $v_{\chi}$ is essential. To estimate this value, we consider the mixing angle between $Z$ and $Z'$ gauge bosons. To achieve this, we first calculate the masses and eigenstates of the hermitian neutral bosons. In addition to the photon mass, $M_\gamma=0$, and its eigenstate $A_\mu=\frac{1}{N_\gamma}(2\sqrt{3}\,B_\mu+\sqrt { 3 }g_L\,g_N\,W_\mu^3-g_N\,W_\mu^8)$ (where $N_\gamma$ is the normalization constant), it is possible to write analytical masses for the other two neutral bosons, $Z_1$ and $Z_2$, as
\begin{eqnarray}
\label{eq:15}
M_{Z_1}^2,\,M_{Z_2}^2&=&\frac{1}{2}\left[M_Z^2+M_{Z'}^2\mp\sqrt{(M_Z^2-M_{Z'}^2)^2+(2M_{ZZ'}^2)^2}\right],
\end{eqnarray} 
where
\begin{eqnarray}
&&M_Z^2=\frac{g_L^2}{4\cos^2\theta_W}\,v_{\text{SM}}^2,\nonumber\\
&&M_{Z'}^2=\frac{g_L^2}{12(1-\frac{1}{3}\tan^2\theta_W)}\bigg[(1+\tan^2\theta_W)^2 v_{\text{SM}}^2-4\tan^2\theta_W v_{\eta}^2+4v_{\chi}^2\bigg],\nonumber\\
&&M_{ZZ'}^2=-\frac{g_L^2}{4\cos^2\theta_W}\frac{v_{\text{SM}}^2-2\cos^2 \theta_W v_{\eta}^2}{\sqrt{3-4\sin^2\theta_W}}\label{eq:16}.
\end{eqnarray}
The angle $\theta_W$ in eq.~\eqref{eq:16} is the Weinberg angle, measured as $\sin^2 \theta_W=0.23121\pm 0.00004$ in the MS-bar scheme \cite{PDG}. The change of basis from $Z_1-Z_2$ to $Z-Z'$ is realized by an SO(2) transformation, with an angle defined as $\theta_{Z}=\frac{1}{2}\tan^{-1}\left(\frac{2M^2_{ZZ'}}{M^2_Z-M^2_{Z'}}\right)$.
To estimate the values of $v_{\eta}$ and $v_{\chi}$, 
we begin by noting that $M_Z$, the mass of the 
SM gauge neutral boson, is measured to be 
$91.1876 \pm 0.0021$ GeV \cite{MZ}. The mixing angle 
$\theta_Z$ is constrained to $-3.98\times10^{-3} 
\lesssim \theta_{Z} \lesssim 1.31\times10^{-4}$~\cite{COGOLLO_2008}. With this information, we identify the allowed region for $v_{\eta}$ and 
$v_{\chi}$, as depicted in Fig.~\ref{fig:veta-vchi}.

Further considerations involve the lower bound on the mass of a new charged gauge boson, set at 6 TeV at a 95\% C.L. according to experimental searches \cite{MWprime1,MWprime2}. In this current context, the mass of the $Y^{\pm}$ boson defined in eq.~\eqref{eq:14} can be approximated as $M^2_{Y^{\pm}} \approx g_L^2\,v_{\chi}^2/4$, assuming $v_{\chi}\gg v_{\eta}$, which is natural in this model. Finally, employing the previously calculated value of $g_L=0.66$ and the approximate $M^2_{Y^{\pm}}$ expression, we can estimate $v_{\chi}\gtrsim$ $18.1$ TeV. Finally, it is noteworthy that for values of $v_{\chi}$ above or equal to $18.1$ TeV, there are no restrictions on the value of $v_{\eta}$. For the sake of concreteness, we consider the value of $v_{\eta}$ that sets $\theta_Z=0$, in other words, the value of $v_{\eta}$ that sets $M^2_{ZZ'}$ to zero in eq. \eqref{eq:16}. Consequently, we set $v_{\eta}\approx 197.5$  GeV, and as a result, $v_{\rho} \approx 147 $ GeV.
\\

In the scalar sector, it is essential to note that the three scalar triplets possess eighteen degrees of freedom. Eight of these correspond to would-be Nambu-Goldstone bosons, which were gauged away to give mass to eight gauge bosons. Thus, we anticipate the presence of ten physical scalar bosons in the mass spectrum. Specifically, the physical scalar spectrum consists of four charged scalars, $H^{\pm}_1,\ H^{\pm}_2$, four CP-even scalars called $h$ (the $125$~GeV SM-like scalar), $H_0$, $H$ and $H'$, and two CP-odd scalars $H_3$ and $H_5$. Some of the analytical expressions for the masses are relatively straightforward to calculate, as outlined below:
\begin{eqnarray}
&& M^2_{H_0}=\frac{v_{\eta}^2+v_{\chi}^2}{2v_{\eta}v_{\chi}}(v_{\rho}\lambda_{15}+v_{\eta}v_{\chi}(\lambda_7+2\lambda_{10})),\label{eq:17}\\
&& M_{H_{1}^{\pm}}^2=\frac{v_{\text{SM}}^2}{2v_{\eta}v_{\rho}}(\lambda_{15}v_{\chi}+\lambda_9v_{\eta}v_{\rho}),\quad
M_{H_{2}^{\pm}}^2=\frac{v_{\rho}^2+v_{\chi}^2}{2v_{\rho}v_{\chi}}(\lambda_{15}v_{\eta}+\lambda_8v_{\rho}v_{\chi}),
\label{eq:18}\\
&& M_{H_3}^2=\frac{v_{\eta}^2+v_{\chi}^2}{2v_{\eta}v_{\chi}}(v_{\rho}\lambda_{15}+v_{\eta}v_{\chi}(\lambda_7- 2\lambda_{10})),\label{eq:19}\\
&& M_{H_5}^2=\frac{\lambda_{15}}{2v_{\eta}v_{\rho}v_{\chi}}(v_{\eta}^2v_{\rho}^2+v_{\text{SM}}^2 v_{\chi}^2).
\label{eq:20}
\end{eqnarray}
In this study, we will assume $\lambda_{10}=0$. It is important to highlight that this condition not only simplifies the model significantly but also induces a mass degeneracy between $H_0$ and $H_3$. Under this assumption, $H_0$ and $H_3$ serve as the real and imaginary components of a complex neutral scalar, respectively.

Concerning the other CP-even scalars ($h$, $H$, and $H'$), obtaining exact analytical expressions for their masses proves to be overly intricate and not particularly practical. However, we can take advantage of the hierarchy $v_\chi\gg v_\rho,\,v_\eta$ to determine approximate analytical expressions by using a perturbative approach \cite{Sakurai}. To achieve this, we express $\lambda_{15}$ as $\kappa \, v_{\chi}$, where $\kappa \ll 1$, considering that $\lambda_{15}$ is a small and dimensionful coupling constant. 

More specifically, these CP-even scalars reside in the subspace with basis $(\text{Re}\,\eta^{0}_1,$ $ \text{Re}\,\rho^{0}_2, \text{Re}\,\chi^{0}_3)$, and their squared masses are determined by the matrix
\begin{equation}
    M^2_{\text{CP-even}}=
    \begin{pmatrix}
        2\lambda_{1}v^2_{\eta}+\kappa\frac{v_{\rho}v_{\chi}^2}{2v_{\eta}} & \lambda_{6}v_{\eta}v_{\rho}-\kappa\frac{v_{\chi}^2}{2} & \lambda_{4}v_{\eta}v_{\chi}-\kappa\frac{v_{\rho}v_{\chi}}{2}\\
        \lambda_{6}v_{\eta}v_{\rho}-\kappa\frac{v_{\chi}^2}{2} & 2\lambda_{2}v_{\rho}^2+\kappa\frac{v_{\eta}v_{\chi}^2}{2v_{\rho}} & \lambda_{5}v_{\rho}v_{\chi}-\kappa\frac{v_{\eta}v_{\chi}}{2}\\
        \lambda_{4}v_{\eta}v_{\chi}-\kappa\frac{v_{\rho}v_{\chi}}{2} & \lambda_{5}v_{\rho}v_{\chi}-\kappa\frac{v_{\eta}v_{\chi}}{2} & 2\lambda_{3}v_{\chi}^2+\kappa\frac{v_{\eta}v_{\rho}}{2}
    \end{pmatrix}.\label{eq:MCP-even}
\end{equation}
The key point is to notice that the exact matrix in eq.~\eqref{eq:MCP-even}, when applied to the VEV hierarchy, can be decomposed into
\begin{equation}
    M_{\text{CP-Even}}^2/v_\chi^2 =M_{0}^{2}+M_{1}^{2}+M^2_{2},
\end{equation}
where
\begin{equation}
     M_{0}^2=
    \begin{pmatrix}
 \frac{\kappa\,v_{\rho}}{2v_{\eta}} & -\frac{\kappa}{2} & 0 \\ \noalign{\smallskip}
 -\frac{\kappa}{2} & \frac{\kappa\,v_{\eta}}{2v_{\rho}} & 0 \\ \noalign{\smallskip}
 0 & 0 & 2\lambda_3
    \end{pmatrix},\quad\quad 
    M_{1}^{2}=
     \begin{pmatrix}
 0 & 0 & \lambda_4\varepsilon_2-\frac{\kappa}{2}\varepsilon_1 \\ \noalign{\smallskip}
 0 & 0 & \lambda_5\varepsilon_1-\frac{\kappa}{2}\varepsilon_2 \\ \noalign{\smallskip}
 \lambda_4\varepsilon_2-\frac{\kappa}{2}\varepsilon_1 & \lambda_5\varepsilon_1-\frac{\kappa}{2}\varepsilon_2 & 0
    \end{pmatrix}\ ,\label{Eq:MP2e01}
\end{equation}
\begin{equation}
    \quad M_{2}^{2}=
    \begin{pmatrix}
 2\lambda_1\varepsilon_2^2  & \lambda_6\varepsilon_1\varepsilon_2 & 0 \\ \noalign{\smallskip}
 \lambda_6\varepsilon_1\varepsilon_2 &  2\lambda_2\varepsilon_1^2 & 0 \\ \noalign{\smallskip}
 0 & 0 & \frac{\kappa}{2}\varepsilon_1\varepsilon_2
    \end{pmatrix},\
\end{equation}
with $\varepsilon_1=\frac{v_{\rho}}{v_{\chi}}\ll 1$ and $\varepsilon_2=\frac{v_{\eta}}{v_{\chi}}\ll 1$.
By straightforwardly applying standard perturbation theory up to the order of $\varepsilon_{1}^2$ and $\varepsilon_{2}^2$, we derive the following analytical expressions for the masses:
\begin{align}
     m^2_{h}\approx &\frac{2}{v^2_{\text{SM}}}(\lambda_{1}v^4_{\eta}+\lambda_2v_{\rho}^4+\lambda_6v_{\eta}^2v_{\rho}^2)-\frac{1}{2\lambda_3 v^2_{\text{SM}}}\left(\lambda_4v_{\eta}^2+\lambda_5v_{\rho}^2-\frac{\lambda_{15}v_{\eta}v_{\rho}}{v_{\chi}}\right)^2,\label{eq:mh}\\
    m^2_{H}\approx\,&\frac{\lambda_{15}v^2_{\text{SM}}v_{\chi}}{2v_{\eta}v_{\rho}}+\frac{2v_\eta^2v_\rho^2}{v^2_{\text{SM}}}\left(\lambda_1+\lambda_2-\lambda_6\right)\notag\\&+\frac{v_\eta v_\rho\left[2\left(\lambda_4-\lambda_5\right)v_\eta v_\rho v_\chi +\lambda_{15}\left(v_\eta^2-v_\rho^2\right)\right]^2}{2v^2_{\text{SM}}v_\chi\left(\lambda_{15}v^2_{\text{SM}}-4\lambda_3v_\eta v_\rho v_\chi\right)},\label{eq:mH}\\
    m^2_{H'}\approx\,&2\lambda_3v^2_{\chi}+\frac{\lambda_{15}v_{\eta}v_{\rho}}{2v_{\chi}}-\frac{\lambda_4^2\lambda_{15}v_\eta^4v_\chi^2}{A}-\frac{\lambda_5^2\lambda_{15}v_\rho^4v_\chi^2}{A}\notag\\&+\frac{v_\eta^3v_\rho v_\chi\left[4\lambda_3\lambda_4^2v_\chi^2+\left(\lambda_3+2\lambda_4\right)\lambda_{15}^2\right]}{A}+\frac{v_\eta v_\rho^3 v_\chi\left[4\lambda_3\lambda_5^2v_\chi^2+\left(\lambda_3+2\lambda_5\right)\lambda_{15}^2\right]}{A}\notag\\&-\frac{\lambda_{15}v_\eta^2v_\rho^2\left[\left(\lambda_4\lambda_5+2\lambda_3\lambda_4+
2\lambda_3\lambda_5\right)2v_\chi^2+\lambda_{15}^2\right]}{A}\label{eq:mH'},
\end{align}
where $A$ is defined as
\begin{equation*}
    A=2\lambda_3v^2_{\chi}(4\lambda_3v_{\eta}v_{\rho}v_{\chi}-\lambda_{15}v^2_{\text{SM}}).
\end{equation*}
These approximate expressions in eqs. \eqref{eq:mh}-\eqref{eq:mH'} exhibit only a 3\% relative error when compared to the exact masses in a numerical analysis, which is sufficient for our purposes. Additionally, note that $m_{H'} > m_{H} > m_{h}$ due to the hierarchy between the vacuum expectation values and $\lambda_3 > \lambda_{15} / v_\chi$.

Another important point is that, at this level of approximation, the SM-like Higgs $h$ and $H$ are given by 
\begin{eqnarray}
&& h\,=\,\cos{\theta}\,(\text{Re}\,\rho^0_2)+\sin{\theta}\,(\text{Re}\,\eta^0_1),\label{eq:28}\\
&& H\,=-\sin{\theta}\, (\text{Re}\,\rho^0_2)+\cos{\theta}\,(\text{Re}\,\eta^0_1),\label{eq:29}
\end{eqnarray}
where $\sin{\theta}=v_{\eta}/v_{\text{SM}}$ and $\cos{\theta}=v_{\rho}/v_{\text{SM}}$. This implies that the coupling between $h$ and the SM leptons, originating from $\text{Y}'_{aa'}\overline{L}_{aL}\rho e_{a'R}$ in eq.~\eqref{eq:8}, can be expressed, after rewriting the symmetry eigenstates $\text{Re}\,\rho^0_2$ and $\text{Re}\,\eta^0_1$ in terms of the mass eigenstates $h$ and $H$, as
\begin{align}
\mathcal{L}^{\rho}&\supset \cos\theta \frac{h}{v_\rho} \left(m_e\overline{e}e+m_\mu\overline{\mu}\mu+m_\tau\overline{\tau}\tau\right)\nonumber \\ \noalign{\smallskip}
&\supset \frac{v_{\text{SM}}\cos\theta}{v_\rho} \frac{h}{v_{\text{SM}}} \left(m_e\overline{e}e+m_\mu\overline{\mu}\mu+m_\tau\overline{\tau}\tau\right)\nonumber \\ \noalign{\smallskip}
&\supset \frac{h}{v_{\text{SM}}} \left(m_e\overline{e}e+m_\mu\overline{\mu}\mu+m_\tau\overline{\tau}\tau\right).
\end{align}
Therefore, we can conclude that the $h$ scalar actually behaves, at this level of the approximation, as the SM Higgs. 

Finally, additional constraints on the $\lambda_{i}$ couplings can be derived from the positivity of the squared scalar masses in eqs.~\eqref{eq:17}-\eqref{eq:20} and from the lower bound obtained in the search for charged scalars which defines $m_{H^{\pm}}>155$~GeV~\cite{MHcharged}. This information allows us to establish lower bounds for $\lambda_{7}$, $\lambda_{8}$ and $\lambda_{9}$,
\begin{eqnarray}
    &&\lambda_{7}\geq -\frac{\lambda_{15}v_{\rho}}{v_{\eta}v_{\chi}},\quad
    \lambda_{8}\geq \frac{2m^2_{H^{\pm}}}{v^2_{\rho}+v^2_{\chi}}-\frac{\lambda_{15}v_{\eta}}{v_{\rho}v_{\chi}},\quad
    \lambda_{9}\geq \frac{2m^2_{H^{\pm}}}{v^2_{\text{SM}}} -\frac{\lambda_{15}v_{\chi}}{v_{\eta}v_{\rho}}.
\end{eqnarray}

Regarding the quark sector, the exact mass matrices are
\begin{eqnarray*}
    && M_{u} = \frac{1}{\sqrt{2}}
    \begin{pmatrix}
 \beta_1v_\eta & \beta_2v_\eta  & \beta_3v_\eta  & 0 \\ \noalign{\smallskip}
 \alpha_{21}v_\rho & \alpha_{22}v_\rho  & \alpha_{23}v_\rho  & 0 \\ \noalign{\smallskip}
 \alpha_{31}v_\rho & \alpha_{32}v_\rho  & \alpha_{33}v_\rho  & 0 \\ \noalign{\smallskip}
 0 & 0 & 0 & \gamma_4v_\chi
    \end{pmatrix},\quad  
   M_{d} = \frac{1}{\sqrt{2}}
     \begin{pmatrix}
 \alpha_1v_\rho & \alpha_2v_\rho  & \alpha_3v_\rho  & 0 & 0 \\ \noalign{\smallskip}
 \beta_{21}v_\eta & \beta_{22}v_\eta  & \beta_{23}v_\eta  & 0 & 0 \\ \noalign{\smallskip}
 \beta_{31}v_\eta & \beta_{32}v_\eta  & \beta_{33}v_\eta  & 0 & 0 \\ \noalign{\smallskip}
 0 & 0 & 0 & \gamma_{24}v_\chi & 0 \\ \noalign{\smallskip}
 0 & 0 & 0 & 0 & \gamma_{35}v_\chi
    \end{pmatrix}.
\end{eqnarray*}
Observe that the $3 \times 3$ sub-matrix pertains to the SM quark sector in both cases, whereas the exotic quarks, within the lower sub-matrix, are diagonal due to $\mathbb{Z}_2$ symmetry. In a general context, determining $\alpha$ and $\beta$ angles involves considering quark masses and their mixings in concordance with experimental data, specifically the measured masses and $V_{\text{CKM}}$ angles. For the sake of completeness, we provide these values below obtained from ref.~\cite{PDG}
\begin{align}
    &m_{u}=2.16\ \substack{+0.49 \\ -0.26}\ \text{MeV},\quad  m_{d}=4.67\  \substack{+0.48 \\ -0.17}\ \text{MeV},\quad m_{s}=93.4\ \substack{+8.6 \\ -3.4}\ \text{MeV},\\
    &m_{c}=1.27 \pm 0.02\ \text{GeV},\ \ 
    m_{b}=4.18 \substack{+0.03 \\ -0.02}\ \text{GeV},\quad\ \,  
    m_{t}=172.69 \pm 0.30\ \text{GeV}\\
    & \theta_{12}=13.00^{\circ} \pm 0.04^{\circ},\,\quad\,\,
    \theta_{23}=2.40^{\circ} \substack{+0.05^{\circ} \\ -0.04^{\circ}},\label{eq:34}\quad\quad\ \ \, 
    \theta_{13}=0.211^{\circ} \pm 0.006^{\circ},\\
    &\delta=65.5^{\circ} \pm 1.5^{\circ}.\label{eq:35}
\end{align}
As evident from eqs.~\eqref{eq:34}-\eqref{eq:35}, the mixing angles are small, implying an approximate equality between mass and symmetry states. Consequently, in this work, we assume that the matrices are diagonal. Notably, the relatively large mass of the top quark justifies the consideration of its Yukawa coupling as the only one significant within the SM quark sector, as it is common practice. A similar scenario happens in the SM lepton sector, i.e. there are no leptonic Yukawa couplings that can significantly modify the renormalization group equations.
\par
After analyzing the mass spectrum of the model, we can now proceed to the study of the stability of the scalar potential at the one-loop level, which we will undertake in next section.

\section{Vacuum stability at one-loop level \label{stability}}

To investigate the stability of the model at the one-loop level, it is important to revisit the tree level analysis, namely the classical vacuum stability. 
In the study presented in ref. \cite{S_nchez_Vega_2019}, classical conditions to guarantee the lower bound of the scalar potential were derived using the method of orbit space along with the copositivity conditions, as detailed in refs.~\cite{kim1982general,kannike2012vacuum, kannike2016vacuum}. 
In summary, this method involves formulating invariants as functions of the fields and their norms, represented by the orbit space parameters $\boldsymbol{\theta}=\frac{f_{ijkl}\phi^{*}_{i}\phi_{j}\phi^{*}_{k}\phi_{l}}{|\phi^{*}_{a}\phi_{a}|^2}$, where a sum over repeated indices is implied. Therefore, all the requisite information to determine the potential minima is encapsulated within these parameters. Consequently, to determine the scalar potential minima for all possible directions in the field space, especially in the large field limit, it suffices to focus on the quartic terms, which is
\begin{eqnarray}
\label{eq:30}
V_4&=&\lambda_1 |\eta |^4 +\lambda_2 | \rho |^4 +\lambda_3| \chi |^4 + (\lambda_4 + \lambda_7 \boldsymbol{\theta}_1 +|\lambda_{10}| \boldsymbol{\theta}_4) |\eta |^2|\chi |^2  \nonumber \\ 
&+& (\lambda_5 +\lambda_8 \boldsymbol{\theta}_2) |\rho |^2 |\chi |^2  + (\lambda_6 +\lambda_9 \boldsymbol{\theta}_3) |\eta |^2|\rho |^2,
\end{eqnarray}
where the orbit space parameters are
\begin{eqnarray}
    && \boldsymbol{\theta}_1(\hat{\eta},\hat{\chi}) =   \hat{\chi}^*_j\hat{\eta}_j\hat{\eta}^*_i\hat{\chi}_i=|\boldsymbol{\theta}_1|, \quad {\boldsymbol{\theta}}_2 (\hat{\rho},\hat{\chi}) =  \hat{\chi}^*_j\hat{\rho}_j\hat{\rho}^*_i\hat{\chi}_i=|\boldsymbol{\theta}_2|,\\ 
    && {\boldsymbol{\theta}}_3 (\hat{\eta},\hat{\rho}) = \hat{\eta}^*_i\hat{\rho}_i \hat{\rho}^*_j\hat{\eta}_j=|\boldsymbol{\theta}_3|,\quad
    \boldsymbol{\theta}_4(\hat{\eta},\hat{\chi}) = 2|\boldsymbol{\theta}_1|\cos{\omega_{\theta_1}}=-2|\boldsymbol{\theta}_1|.
\end{eqnarray}
Note that for $\boldsymbol{\theta}_4$ we used $\omega_{\theta_1}=\pi$, which is the correct value for the vacuum.
One can think of the $\boldsymbol{\theta}$ frontier as a cube with edge length equal to one. However, as pointed out in~\cite{S_nchez_Vega_2019}, the $\boldsymbol{\theta}$ frontier is given by
\begin{align}
    &\resizebox{.25\hsize}{!}{$0 \leq |\boldsymbol{\theta}_{1}|\leq 1, \quad  0 \leq |\boldsymbol{\theta}_{2}|\leq 1,$} \nonumber \\ \noalign{\smallskip}
    &\resizebox{0.8\hsize}{!}{$\max \left[0,\,\sqrt{|\boldsymbol{\theta}_{1}||\boldsymbol{\theta}_{2}|}-\sqrt{\left( 1-|\boldsymbol{\theta}_{1}|\right)\left( 1-|\boldsymbol{\theta}_{2}|\right)}  \right]^2 \leq |\boldsymbol{\theta}_{3}| \leq   \left[\sqrt{|\boldsymbol{\theta}_{1}||\boldsymbol{\theta}_{2}|}+\sqrt{\left( 1-|\boldsymbol{\theta}_{1}|\right)\left( 1-|\boldsymbol{\theta}_{2}|\right)}  \right]^2.$}
\end{align}
\par
Notice that $V_4$ can be expressed in a biquadratic form involving the norms of the fields, represented as $\mathbf{h}^\text{T}\Lambda\mathbf{h}$, where $\mathbf{h}^{\text{T}}=(|\eta|^2,\, |\rho|^2,\, |\chi|^2)\geq0$ and, the matrix $\Lambda$ is
\begin{eqnarray}
\label{eq:27}
\Lambda =
\begin{pmatrix}
\lambda_1 & (\lambda_6 +\lambda_9 \boldsymbol{\theta}_3)/2  & 
(\lambda_4 + \lambda_7 \boldsymbol{\theta}_1 -2\lambda_{10} \boldsymbol{\theta}_1) /2\\
\star &\lambda_2 & (\lambda_5 +\lambda_8 \boldsymbol{\theta}_2)/2\\
\star & \star & \lambda_3
\end{pmatrix}.
\end{eqnarray}
Therefore, the scalar potential is bounded from below, if the symmetric matrix $\Lambda$ is strictly copositive. The use of strong stability requirement, $V_4>0$, is essential to accommodate the $\lambda_{15}$ term, as the requirement in the marginal sense, $V\geq 0$, forbids cubic terms. 

By applying the copositivity conditions to the $\Lambda$ matrix\footnote{The completed calculation and details can be see in~\cite{S_nchez_Vega_2019}.}, one determines seventeen inequalities that serve to constrain the allowed values of $\lambda_{i}$ couplings,
\begin{eqnarray}
& & \lambda_1 > 0,\quad \lambda_2 >0,\quad \lambda_3 >0,\label{eq:41}\\
& &\lambda_4 +2\sqrt{\lambda_1 \lambda_3}>0, \quad \lambda_4 + \lambda_7  -2 |\lambda_{10}|+2\sqrt{\lambda_1 \lambda_3}>0, \\
& &\lambda_5 + 2\sqrt{\lambda_2 \lambda_3}>0,\quad \lambda_5 +\lambda_8 + 2\sqrt{\lambda_2 \lambda_3}>0, \\
& &\lambda_6 + 2\sqrt{\lambda_1 \lambda_2}>0, \quad 
\lambda_6 +\lambda_9 + 2\sqrt{\lambda_1 \lambda_2}>0, \\
& & \textrm{C}_{1}\sqrt{\lambda_2}+\textrm{C}_{2}\sqrt{\lambda_1}+\textrm{C}_{3}\sqrt{\lambda_3}+2\sqrt{\lambda_1 \lambda_2 \lambda_3}+\sqrt{\overline{\textrm{C}}_{1}\overline{\textrm{C}}_{2}\overline{\textrm{C}}_{3}}> 0\label{eq:45},
\end{eqnarray}
where $ \textrm{C}_1=\lbrace \lambda_4,\,\lambda_4 + \lambda_7 -2 |\lambda_{10}|\rbrace$, $ \textrm{C}_2=\lbrace  \lambda_5,\,\lambda_5 +\lambda_8 \rbrace$, $ \textrm{C}_3= \lbrace  \lambda_6,\,\lambda_6 +\lambda_9 \rbrace$, $\overline{\textrm{C}}_1= \textrm{C}_1+2\sqrt{\lambda_1\lambda_{3}}$, $ \overline{\textrm{C}}_2=\textrm{C}_2+2\sqrt{\lambda_2\lambda_{3}}$ and $\overline{\textrm{C}}_3=\textrm{C}_3+2\sqrt{\lambda_1\lambda_{2}}$.

Now, our attention shifts to the examination of model stability at the one-loop level. It is well-established that, at this order, numerous interactions not present in the original Lagrangian may manifest, giving rise to substantial modifications in the stability characteristics of the model. Not only that, but at one-loop level, the parameters acquire an energy scale-dependence, as a consequence of the renormalization procedure, and the parameters are referred to as running couplings. This procedure generates first order differential equations known as Renormalization Group Equations, or RGE, for short. It could tell us how the parameters change, when a change in scale is made.

For models like the SM and extensions, such as 331 models, it is not an easy task to determine and solve the RGE, due to the large number of Feynman diagrams associated to the large number of parameters. Fortunately, there are numerous softwares to compute these one loop diagrams and the RGE, see for instance ref.~\cite{thomsen2021introducing,staub2012sarah,Sartore_2021}. In our analysis, we have used RGBeta \cite{thomsen2021introducing}. This package can provide calculations up to four-loop order for the gauge couplings and up to the three-loop order for the remaining parameters. Nevertheless, for the current work, a calculation at the one-loop level is more than sufficient to illustrate the primary behavior of the model at the quantum level. 

The renormalization group equations take on a generic form at the one-loop level, expressed as $\frac{d{X}}{dt}\equiv\beta_{X}/(4\pi)^2$ (where $t$ is a real parameter defined as $t=\log{\mu/\mu_0}$, and in this work, we use $\mu_0=m_Z$), with $X$ representing a generic coupling. Using RGBeta, we derive the following expressions for the $\beta$-functions: 
\begin{equation}
    \beta_{g_N}=
    \frac{26}{3}g_N^3,
    \ \beta_{g_L}=-\frac{13}{2}g_L^3,\ \beta_{g_3}=-5g_3^3,
\end{equation}
\begin{equation}
    \beta_{\alpha_{33}}=\alpha_{33}\left(-\frac{4}{3}g_{N}^2-4g_{L}^2-8g_3^2+5\alpha_{33}^2+\frac{1}{2}\gamma_{35}^2\right),
\end{equation}
\begin{equation}
    \beta_{\gamma_4}=\gamma_{4}\left(-\frac{5}{3}g_{N}^2-4g_{L}^2-8g_3^2+5\gamma_{4}^2+3\gamma_{24}^2+3\gamma_{35}^2\right)
\end{equation}
\begin{equation}
    \beta_{\gamma_{24}}=\gamma_{24}\left(-\frac{1}{3}g_{N}^2-4g_{L}^2-8g_3^2+3\gamma_{4}^2+5\gamma_{24}^2+3\gamma_{35}^2\right),
\end{equation} 
\begin{equation}
    \beta_{\gamma_{35}}=\gamma_{35}\left(-\frac{1}{3}g_{N}^2-4g_{L}^2-8g_3^2+\frac{1}{2}\alpha_{33}^2+3\gamma_{4}^2+3\gamma_{24}^2+5\gamma_{35}^2\right),
\end{equation}
\begin{align}
    \beta_{\lambda_1}=&\frac{2}{27}g_{N}^4+\frac{4}{9}g_{N}^2g_{L}^2+\frac{13}{6}g_{L}^4-\left(\frac{4}{3}g_{N}^2+16g_{L}^2\right)\lambda_1+28\lambda_1^2+4\lambda_{10}^2\notag\\
    &+3\lambda_4^2+3\lambda_6^2+2\lambda_4\lambda_7+\lambda_7^2+2\lambda_6\lambda_9+\lambda_9^2,
\end{align}
\begin{align}    
    \beta_{\lambda_2}=&\frac{32}{27}g_{N}^4+\frac{16}{9}g_{N}^2g_{L}^2+\frac{13}{6}g_{L}^4-\left(\frac{16}{3}g_{N}^2+16g_{L}^2\right)\lambda_2+28\lambda_2^2+3\lambda_5^2\notag\\
    &+3\lambda_6^2+2\lambda_5\lambda_8+\lambda_8^2+2\lambda_6\lambda_9+\lambda_9^2-6\alpha_{33}^4+12\alpha_{33}^2\lambda_2,
    \end{align}
\begin{align}
    \beta_{\lambda_3}=&\frac{2}{27}g_{N}^4+\frac{4}{9}g_{N}^2g_{L}^2+\frac{13}{6}g_{L}^4-\left(\frac{4}{3}g_{N}^2+16g_{L}^2\right)\lambda_3+28\lambda_3^2+3\lambda_4^2\notag\\
    &+3\lambda_5^2+2\lambda_4\lambda_7+\lambda_7^2+2\lambda_5\lambda_8+\lambda_8^2-6\gamma_{4}^4-6\gamma_{24}^4-6\gamma_{35}^4\notag\\
    &+12\left(\gamma_{4}^2+\gamma_{24}^2+\gamma_{35}^2\right)\lambda_3+4\lambda_{10}^2,\label{eq:53}
\end{align}
\begin{align}
    \beta_{\lambda_4}=&\frac{4}{27}g_{N}^4-\frac{4}{9}g_{N}^2g_{L}^2+\frac{11}{6}g_{L}^4-\left(\frac{4}
    {3}g_{N}^2+16g_{L}^2\right)\lambda_4+8\lambda_{10}^2+16\lambda_1\lambda_4\notag\\   &+16\lambda_3\lambda_4+4\lambda_4^2+6\lambda_5\lambda_6+4\lambda_1\lambda_7+4\lambda_3\lambda_7+2\lambda_7^2+2\lambda_6\lambda_8\notag\\
    &+2\lambda_5\lambda_9+6\left(\gamma_{4}^2+\gamma_{24}^2+\gamma_{35}^2\right)\lambda_4,
\end{align}
\begin{align}
    \beta_{\lambda_5}=&\frac{16}{27}g_{N}^4+\frac{8}{9}g_{N}^2g_{L}^2+\frac{11}{6}g_{L}^4-\left(\frac{10}{3}g_{N}^2+16g_{L}^2\right)\lambda_5+\lambda_2\lambda_5+16\lambda_3\lambda_5\notag\\
    &+4\lambda_5^2+6\lambda_4\lambda_6+2\lambda_6\lambda_7+4\lambda_2\lambda_8+4\lambda_3\lambda_8+2\lambda_8^2+2\lambda_4\lambda_9\notag\\
    &+6\left(\alpha_{33}^2+\gamma_{4}^2+\gamma_{24}^2+\gamma_{35}^2\right)\lambda_5,
\end{align}
\begin{align}
    \beta_{\lambda_6}=&\frac{16}{27}g_{N}^4+\frac{8}{9}g_{N}^2g_{L}^2+\frac{11}{6}g_{L}^4-\left(\frac{10}{3}g_{N}^2+16g_{L}^2\right)\lambda_6+6\lambda_4\lambda_5+16\lambda_1\lambda_6\notag\\
    &+16\lambda_2\lambda_6+4\lambda_6^2+2\lambda_5\lambda_7+2\lambda_4\lambda_8+4\lambda_1\lambda_9+4\lambda_2\lambda_9+2\lambda_9^2+6\alpha_{33}^2\lambda_6,
\end{align}
\begin{align}
    \beta_{\lambda_7}=&\frac{4}{3}g_{N}^2g_{L}^2+\frac{5}{2}g_{L}^4-\left(\frac{4}{2}g_{N}^2+16g_{L}^2\right)\lambda_7+40\lambda_{10}^+4\lambda_1\lambda_7+4\lambda_3\lambda_7\notag\\
    &+8\lambda_4\lambda_7+6\lambda_7^2+2\lambda_8\lambda_9+6\left(\gamma_{4}^2+\gamma_{24}^2+\gamma_{35}^2\right)\lambda_7,
\end{align}
\begin{align}
    \beta_{\lambda_8}=&-\frac{8}{3}g_{N}^2g_{L}^2+\frac{5}{2}g_{L}^4-\left(\frac{10}{3}g_{N}^2+16g_{L}^2\right)\lambda_8+4\lambda_2\lambda_8+4\lambda_3\lambda_8+8\lambda_5\lambda_8\notag\\
    &+6\lambda_8^2+2\lambda_7\lambda_9+6\left(\alpha_{33}^2+\gamma_{4}^2+\gamma_{24}^2+\gamma_{35}^2\right)\lambda_8-12\gamma_{35}^2\alpha_{33}^2,
\end{align}
\begin{align}
    \beta_{\lambda_9}=&-\frac{8}{3}g_{N}^2g_{L}^2+\frac{5}{2}g_{L}^2-\left(\frac{10}{3}g_{N}^2+16g_{L}^2\right)\lambda_9+2\lambda_7\lambda_8+4\lambda_1\lambda_9+4\lambda_2\lambda_9\notag\\
    &+8\lambda_6\lambda_9+6\lambda_9^2+6\alpha_{33}^2\lambda_9,
\end{align}
\begin{align}
    \beta_{\lambda_{10}}=&\lambda_{10}\left[-\frac{4}{3}g_{N}^2-16g_{L}^2+4\lambda_3+8\lambda_4+4\lambda_1+16\lambda_7+6\left(\gamma_{4}^2+\gamma_{24}^2+\gamma_{35}^2\right)\right].\label{eq:48}
\end{align}
\begin{equation}
    \beta_{\lambda_{15}}=\lambda_{15}[-2g^2_{N}-12g^2_L+3(\gamma^2_{24}+\gamma^2_{25}+\gamma_{4}^2+\alpha_{33}^2)+2(\lambda_{4}+\lambda_5+\lambda_6-\lambda_7-\lambda_8-\lambda_9)]
\end{equation}
As discussed in the previous section, when we adopt $\lambda_{10}=0$, a new symmetry emerges. This can also be seen from eq.~\eqref{eq:48}, which vanishes for this choice. Notice that a similar situation occurs for $\lambda_{15}$, which is a U$(1)_{\text{PQ}}$ symmetry studied in \cite{Sanchez-Vega:2016dwe}, of which the well-known $\ZZ_2$ symmetry is a discrete subgroup.

Now, we focus on the issue of the loss of perturbativity in the effective potential at the one-loop level. To do this, let us recall that the effective potential contains logarithmic terms of the ratios of the mass eigenvalues to the renormalization scale. In other words, the effective potential at one-loop level is generically written as:
\begin{equation}
    V_{\text{eff}}(\mu,\,\lambda,\,\phi)=V(\lambda,\,\phi)+\frac{\hbar}{64\pi^2}\sum_{i}n_{i}m^4_{i}(\lambda,\,\phi)\left(\ln{\frac{m^2_{i}(\lambda,\,\phi)}{\mu^2}}+\zeta_{i}\right),\label{Veff1loop}
\end{equation}
where $\hbar$ indicates the one-loop level. The $i$ index runs over all particle species. The $n_i$ is the number of degrees of freedom associated to the field, and $m_i$ is its mass eigenvalue. In addition, $\zeta_{i}$ is a constant that depends on the renormalization scheme, having different values depending on the nature of the field. Moreover, we are considering $\lambda$ and $\phi$ as a collection of quartic couplings and fields, respectively.

In order to maintain perturbativity, the mass scales can not be very different when the energy scale $\mu$ increases, since the large logarithms would spoil perturbativity. However, as we have seen in the previous section, the masses of the SM sector have very distinct values to those of the new particles arising from the economical 331 model. This may create a significant problem related to the perturbativity in the current case. For this reason, we use the method in ref.~\cite{chataignier2018single} which improves the effective potential using a single renormalization scale. 
The key point of this method is to choose a renormalization scale $\mu=\mu^{*}$ where $\mu^{*}$ is on the hypersurface where the one-loop contributions to the effective potential vanishes, that is, that the effective potential in eq.~\eqref{Veff1loop} is equal to the classical potential evaluated in the running coupling $\lambda(\mu^{*})$,
\begin{equation}
    V_{\text{eff.}}(\mu^{*},\,\lambda,\,\phi)=V(\lambda(\mu^{*}),\,\phi(\mu^{*})).
\end{equation}

This choice is motivated by the fact that, while it is not possible to individually suppress all logarithms with a single renormalization scale, it is possible to suppress them altogether when evaluating on this hypersurface. One important consequence of this method is the ability to use classical stability conditions to the improved effective potential. In practice, by solving the RGEs, we can substitute the $\lambda$ on the tree-level stability criteria, eqs.~\eqref{eq:41}-\eqref{eq:45}, by their respective running couplings $\lambda(\mu^{*})$.

With everything settled, our goal is to identify constraints on the masses of the new particles introduced by the model. Specifically, we choose to show bounds as functions of the exotic quark mass and the heaviest scalar, $m_{H'}$, since these particles gain masses predominantly at the 331 scale. Additionally, we aim to uncover the allowed values among the quartic couplings in the parameter space that ensure the stability of the scalar potential at high energies. In order to achieve this, one needs to solve the eighteen RGEs, which are non-linear and coupled, considering a set of initial conditions suitable for the model parameters, which are capable of evolving the entire parameter space up to the Planck scale  $\approx 10^{19}$ GeV. Certain parameters of the economical 331 model have already been determined in Sec. \ref{mass} through additional constraints. These constraints include, for example, ensuring the positivity of squared masses for the scalars, considering the $Z-Z'$ mixing, and taking into account searches for new charged gauge bosons and charged scalars. However, other initial parameters, especially the scalar couplings $\lambda$s, are less restricted and need to be chosen through trial and error, ensuring that they at least satisfy the constraints in eqs.~\eqref{eq:41}-\eqref{eq:45}. A set of parameters that satisfies all the aforementioned constraints and allows the economical 331 model to have stable regions up to the Planck scale is as follows:
\begin{eqnarray}
    && g_{N}=0.374,\quad\quad\ \ \, g_{L}=0.664,\quad\ \,\, g_{3}=1.22,\\
    && v_{\eta}=197.52\ \text{GeV},\ v_{\rho}=147\ \text{GeV},\ v_{\chi}=18.1\ \text{TeV},\\
    && \lambda_1=0.1,\ \lambda_2=0.15,\ \lambda_4=0.032,\ \lambda_5=-0.055,\ \lambda_7=0,\\
    && \lambda_8=0,\  \lambda_9=-0.5,\ \lambda_{10}=0,\ \lambda_{15}=25\ \text{GeV},\ \alpha_{33}=0.5,
\end{eqnarray}
where $g_3=\sqrt{4\pi\alpha_s}$, with $\alpha_s=0.1180 \pm 0.009$ \cite{PDG}. Furthermore, by using the relation $\tan{\theta_W}=-\sqrt{3}\sin{\theta_X}$, where $\theta_X$ is the rotation angle associated to the first symmetry breaking, we find $\theta_X=-18^{\circ}$. With this value, we are able to use the relation $e=g_{N}\cos{\theta_X}\cos{\theta_W}$, where $e$ is the elementary charge associated with the U$(1)_{Q}$ symmetry, in order to obtain $g_{N}=0.374$.

\begin{figure}[h!]
    \includegraphics[scale=0.35]{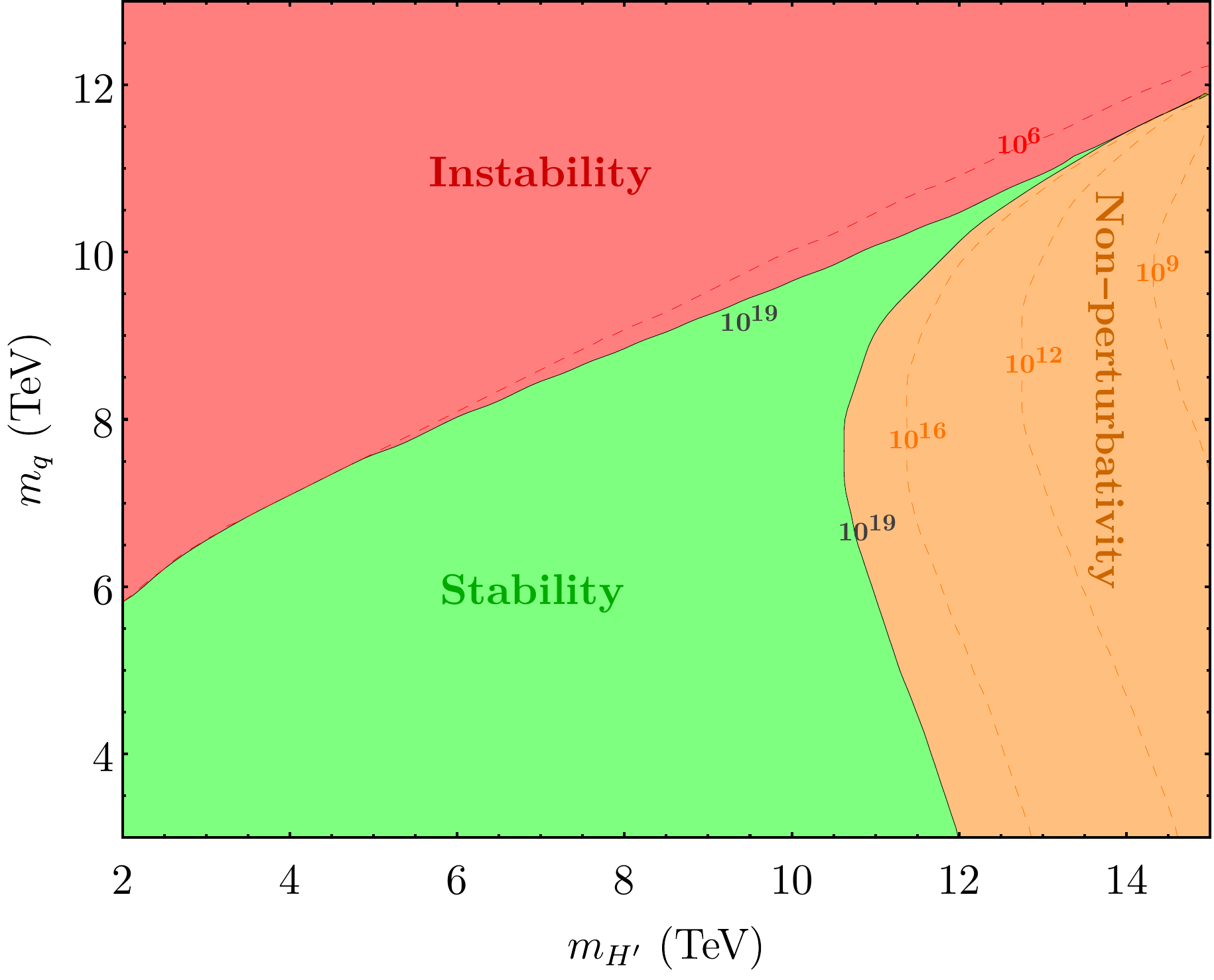}
    
    \caption{\label{fig:mq-mhp plane}The economical 331 model vacuum stability for the allowed values of $m_{H'}$ and $m_{q}$. The non-perturbativity region is defined by values for which $\lambda_{i}>4\pi$. The border (black line) between the three regions occurs on the Planck scale $10^{19}$ GeV. In contrast to the non-perturbative scenario, the stability-instability border exhibits a more rapid convergence of scales. Notably, the non-perturbative boundary diminishes with decreasing scale.}
\end{figure}

The exotic Yukawa couplings (which we assume to be equal in this work because we have no a priori reason to make the exotic quark masses, $m_{q_i}$, different) and $\lambda_3$ are variables in this analysis, since we use them to change the exotic quark mass and the heaviest scalar, $m_{H'}$. As a consequence of this parametrization, we choose the $\lambda_6$ coupling to become a function of $\lambda_{3}$ in order to fix the Higgs mass in eq.~\eqref{eq:mh} to be the measured value. Hence, by applying the tree-level stability criteria, from eqs.~\eqref{eq:41}-\eqref{eq:45}, up to Planck scale $\approx 10^{19}$~GeV, we obtain the region in the $m_{q}-m_{H'}$ plane, shown in Fig. \ref{fig:mq-mhp plane}.

To understand the boundaries shown in Fig. \ref{fig:mq-mhp plane}, we need to resort to the RGEs, more specifically to the $\beta_{\lambda_3}$ function in eq.~\eqref{eq:53}. There are negative contributions coming from the quartic terms of the exotic Yukawa quark couplings, i.e. $\gamma^4_4$, $\gamma^4_{24}$, and $\gamma^4_{35}$ terms, and positive contributions coming from $\lambda_3$ and $\lambda^2_3$ terms. As $m_{H'}$ increases, $\lambda_3$ also increases due to its natural positive contribution. However, as the exotic Yukawa quark couplings increase, $m_{q}$ increases, leading to a decrease in the $\beta_{\lambda_3}$ function.

\begin{figure}[h!]
    \includegraphics[scale=0.33]{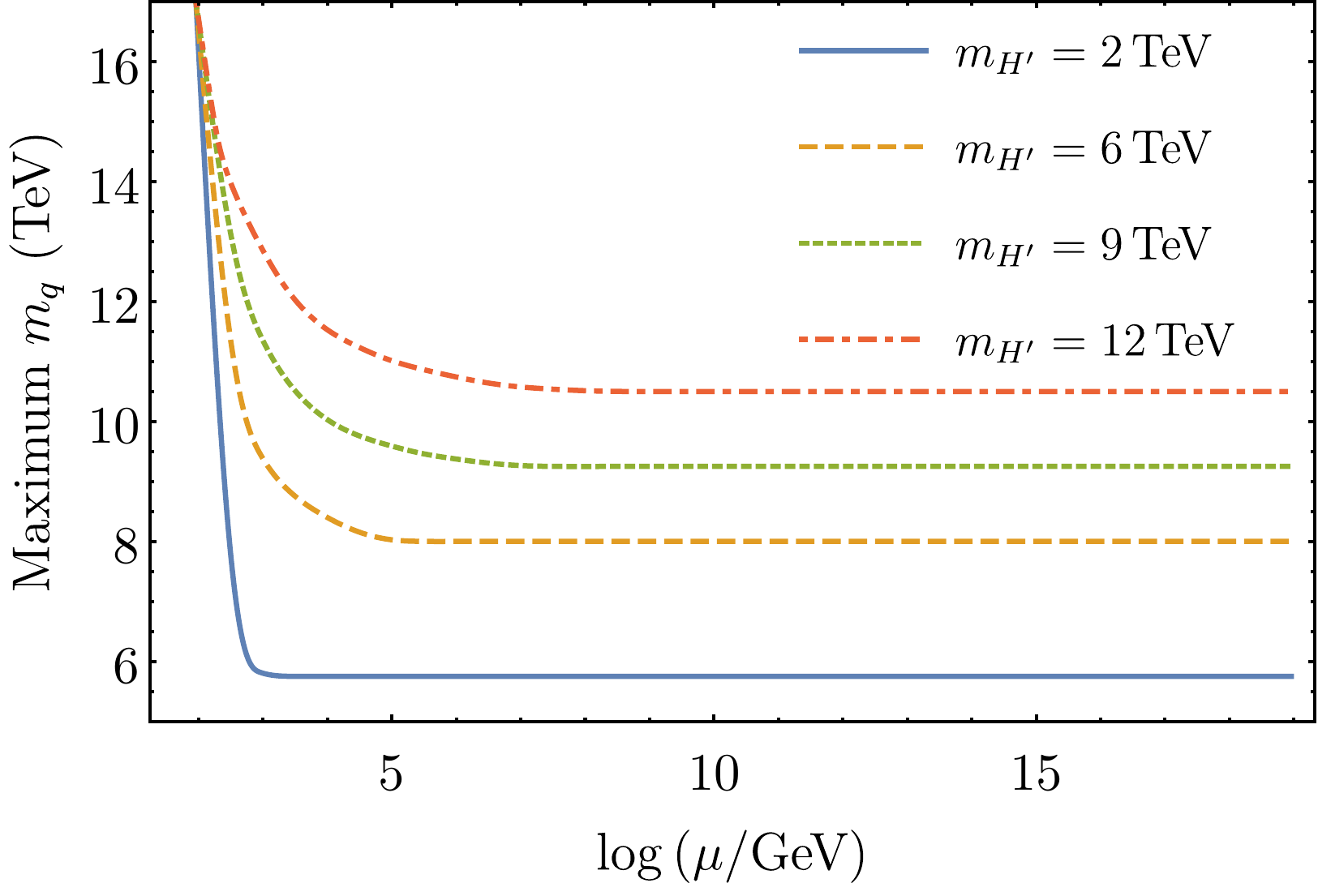}
    \caption{\label{fig:mq-function}The maximum mass value for the exotic quark masses $m_q$ as function of $m_{H'}$ and the energy scale $\mu$. Note that as 
$\mu$ approaches to $m_Z$, which is the renormalization subtraction point, $m_{q}$ has arbitrarily large values. This is to be expected, since the vacuum is classically stable regardless of $m_{q}$. Hence, as the effective couplings begin to evolve, so does $m_{q}$, eventually approaching a stable value.}
\end{figure}

This intricate interplay between the couplings persists until the scalar potential becomes unstable, reaching a maximum value of approximately 14 TeV for $m_{H'}$ and 11 TeV for $m_{q}$. The relation between these couplings is further illustrated in Fig. \ref{fig:mq-function}, where the behaviour of $m_q$ as function of $\log(\mu/\text{GeV})$ is presented. It is clear that for different values of $m_{H'}$, $m_{q}$ reaches different maximum values. For instance, when $m_{H'}=12$ TeV, the maximum value of $m_q$ is approximately 11 TeV, maintaining it constant for energy scales $\gtrsim 10^8$ GeV, a detail that may not be immediately apparent from Fig. \ref{fig:mq-mhp plane}. 

\begin{figure}
    \centering
    \begin{subfigure}{}
    \includegraphics[width=0.4\textwidth]{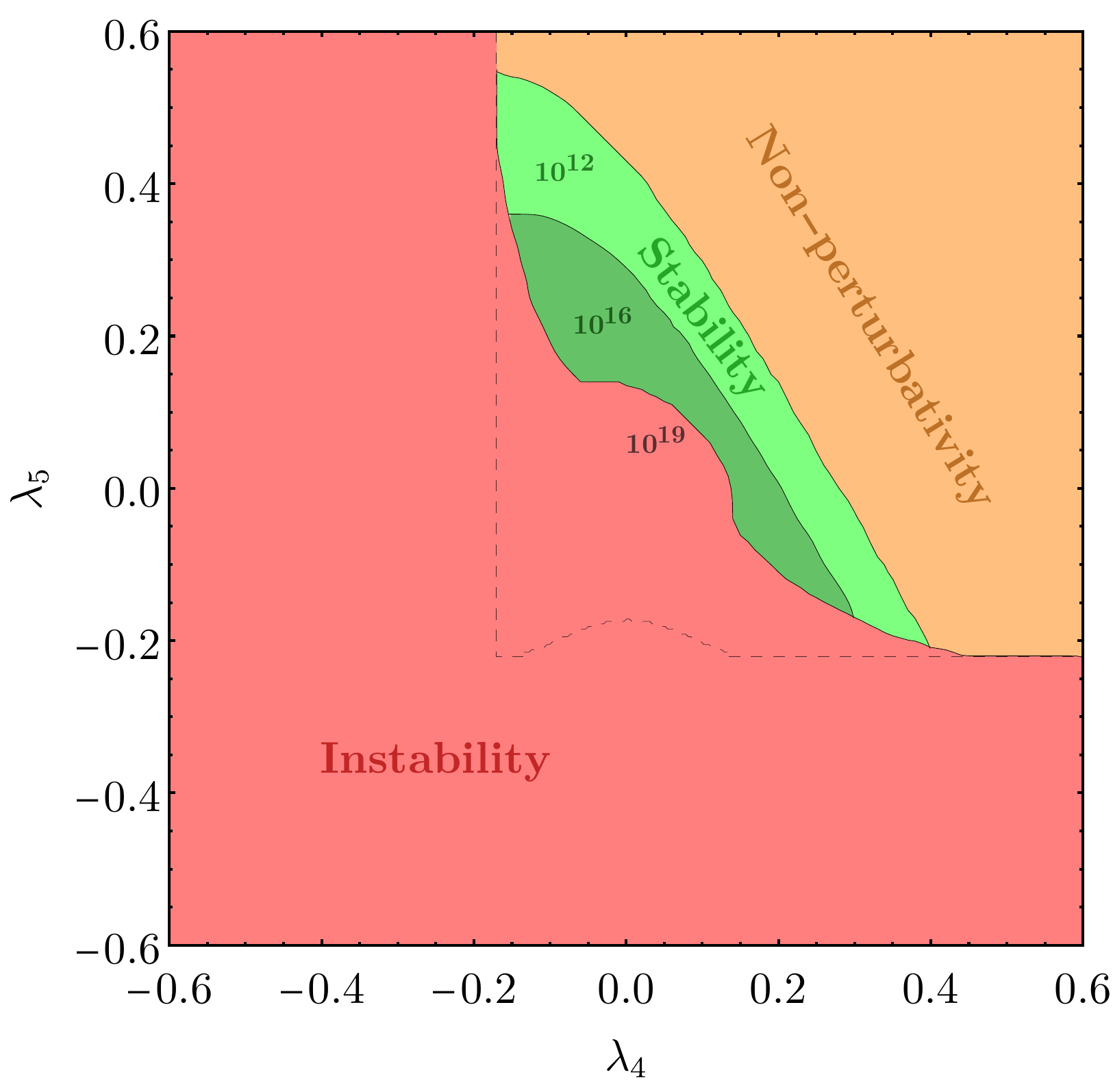}
    \end{subfigure}
    \hspace{1cm}
    \begin{subfigure}{}   \includegraphics[width=0.38\textwidth]{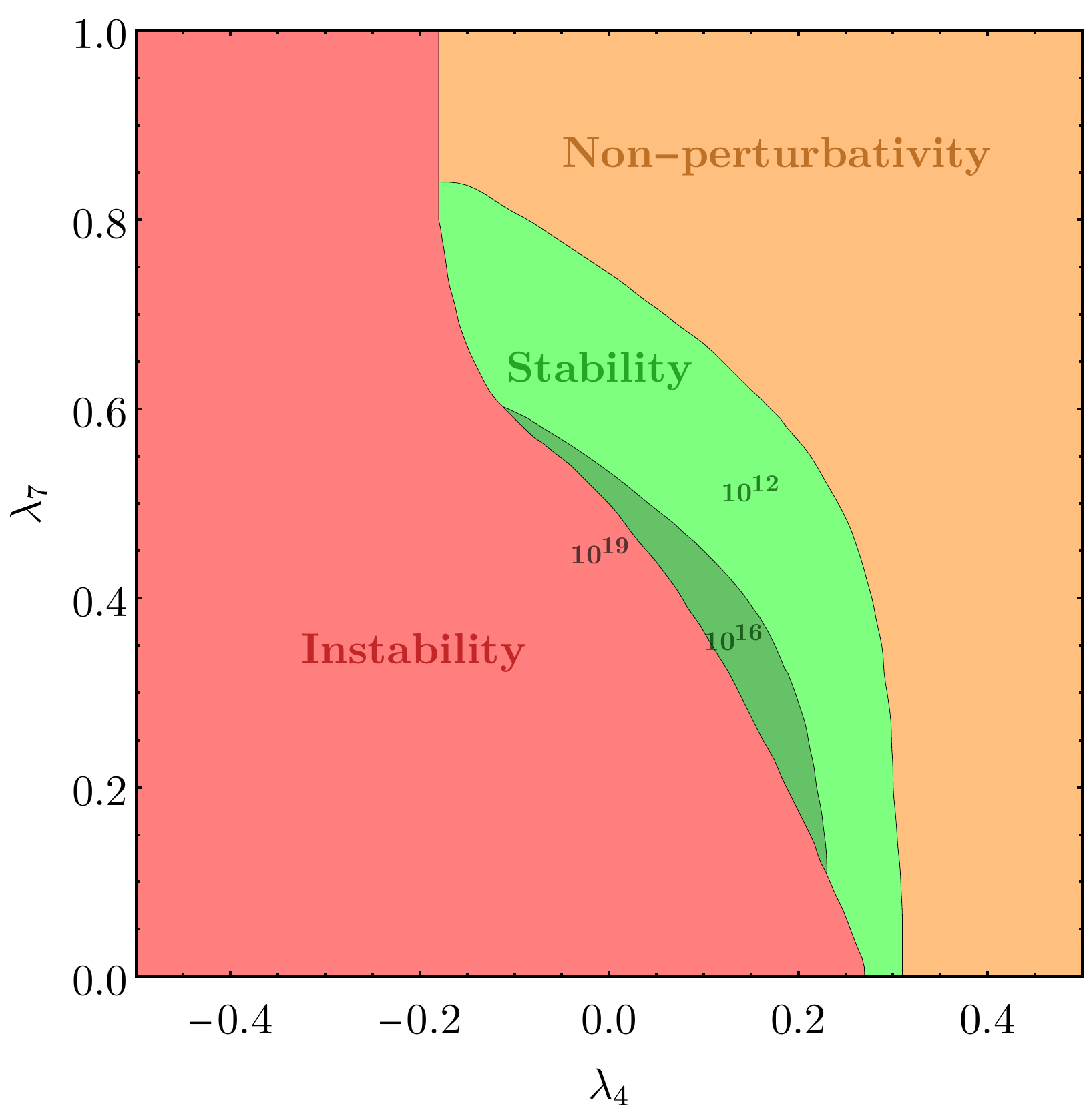}
    \end{subfigure}
    \caption{\label{fig:L4-L7}The dashed line represents the classical region frontier. The lighter green region depicts the allowed values to the potential be stable up to $10^{12}$ GeV, while the darker green one the potential is stable up to $10^{16}$~GeV. Note that the stability up to Planck scale, $10^{19}$~GeV, is not possible within this scenario.}
\end{figure}

Another way to utilize vacuum stability and ensure perturbativity up to the Planck scale to constrain the model is to directly examine the couplings. This approach is particularly relevant as one of these parameters can reflect the intensity of a portal between the SM sector and the new 331 sector. An example of this is observed in dark matter studies through the Higgs portal within the real singlet scalar model \cite{Profumo_2010,Guo_2010,Feng_2015}. With this motivation in mind, we conducted an analysis of the parameter space defined by $\lambda_{4}-\lambda_{5}$ and $\lambda_{4}-\lambda_{7}$ to impose additional constraints on the allowed region within this domain. In our studies, we observed that if the masses of exotic quarks are below $9$ TeV, the one-loop region depicted is nearly identical to the classical one. Therefore, for this particular analysis, we set $m_{q}=9$ TeV. Furthermore, in this scenario, the dependence of $m_h$ on $\lambda_4$ demands the variation of $\lambda_6$, similar to the approach adopted previously for Fig.~\ref{fig:mq-mhp plane}. Consequently, stability is not observed at the Planck scale but only for energy values below $10^{16}$~GeV. Another interesting feature of our analysis is the constraint imposed on $\lambda_7$. In a prior study in ref.~\cite{S_nchez_Vega_2019}, the authors established the maximum and minimum allowed values on the $\lambda_4-\lambda_7$ plane in the classical case. Our findings indicate that, to uphold the positivity of mass square at the one-loop level, $\lambda_7$ must be greater than or equal to zero, otherwise, $M^2_{H_0}$ and $M^2_{H_3}$ assume a negative value.  It is also interesting to compare the vacuum stability region of the economical 331 model at the one-loop level to the classical region obtained in ref. \cite{S_nchez_Vega_2019}. Fig. \ref{fig:L4-L7} clearly shows a significant reduction in the vacuum stability region in the quantum case.


\section{Conclusions\label{conclusions}}
In this work, we study vacuum stability at one-loop of the economical 331 model by resumming the renormalization group equations (RGE) into the tree level potential which are obtained from the \textit{Mathematica} package RGEBeta~\cite{thomsen2021introducing}. We apply the method of single scale renormalization for multiscale effective potential developed in~\cite{chataignier2018single}, allowing us to integrate the RGE and substitute the solutions into the tree level stability criteria determined in ref.~\cite{S_nchez_Vega_2019}. The method allows any boundaries conditions, as long as the one loop potential vanishes on it. Therefore, one may choose arbitrary conditions for the couplings and vary them on a grid, requiring stability up to any scale. For our purposes, the Planck scale is more than enough, in order to compare to the SM or its extensions. The latest fit data for new Higgs-like bosons, charged scalars, charged and neutral gauge bosons are used to set constraints on the model parameters, namely, the symmetry breaking scale $v_{\chi}$, for which we determine a minimum value of $\approx 18.1$ TeV. We also assume a zero mixing angle between $Z-Z'$ allowing us to determine the values of $v_{\eta},\ v_{\rho}$.

In the scalar sector, we assumed a global U(1) symmetry emerging when setting the $\lambda_{10}$ coupling to zero. Similar observations have also been noted in the two Higgs doublet model \cite{Jueid_2020,Ferreira:2020ana}. For the CP-even scalars, we derived more precise expressions for the scalar masses than those shown in~\cite{S_nchez_Vega_2019}. These expressions were obtained by applying traditional non-degenerate perturbation theory, and their formulations align with those in~\cite{cao2016collider} for the first few terms.

We select an appropriate set of parameters for the $\lambda_i$ couplings, perform numerical solutions of the RGE up to the Planck scale, and thereby unravel, for the first time, intricate relations between the mass of the heaviest scalar, $m_{H'}$, and the masses of the exotic quarks, $m_q$, ensuring the stability of the economical 331 model up to the Planck scale, shown in Fig. \ref{fig:mq-mhp plane}. These regions looks similar to the one shown in~\cite{Degrassi_2012}, depicting Higgs-top mass plane in the SM context. Remarkably, a very interesting limit for the heaviest exotic quark mass appears in Fig.~\ref{fig:mq-mhp plane}, and more detailed in Fig.~\ref{fig:mq-function}. 
The first figure illustrates the maximum values of 
$m_{q},\, m_{H'}$ where the model remains stable and perturbative. The latter figure shows an interesting phenomenon: the maximum value for the exotic quark mass stabilizes way before the Planck scale. A similar situation occurs in the analysis of the quark top mass in ref.~\cite{Degrassi_2012}, stabilizing near $10^{12}$ GeV, whereas in our case the upper limit on the heaviest exotic quark stabilizes near $10^{8}$~GeV. Hence, we have determined an upper bound for the heaviest quark mass within the model, which is not so distant from future LHC runs, serving as bounds to be searched.

Finally, we explore interesting relations between the $\lambda_i$ couplings emerging from the vacuum stability and the perturbativity conditions. In particular, we choose to illustrate the different regions of stability in the $\lambda_4-\lambda_5$ and $\lambda_4-\lambda_7$ planes, see Fig.~\ref{fig:L4-L7}. Although, at a first glance, these $\lambda_i$ parameters are not directly related to a physical observable, the allowed region from this figure is useful to constrain the large set of parameters of the economical 331 model. 


\begin{acknowledgements}
\noindent
{B. L. S\'anchez-Vega thanks the National Council for Scientific and Technological Development of Brazil, CNPq, for the financial support through grant n$^{\circ}$ 311699/2020-0. A. A. Louzi thanks the Coordination for the Improvement of Higher Educational Personnel, CAPES, for the financial support.}
\end{acknowledgements}


\providecommand{\href}[2]{#2}\begingroup\raggedright
\endgroup

\end{document}